\documentclass[12pt]{article}
\begin{document}
\newcommand{\eqn}[1]{eq.(\ref{#1})}
\renewcommand{\section}[1]{\addtocounter{section}{1}
\vspace{5mm} \par \noindent
  {\bf \thesection . #1}\setcounter{subsection}{0}
  \par
   \vspace{2mm} } 
\newcommand{\sectionsub}[1]{\addtocounter{section}{1}
\vspace{5mm} \par \noindent
  {\bf \thesection . #1}\setcounter{subsection}{0}\par}
\renewcommand{\subsection}[1]{\addtocounter{subsection}{1}
\vspace{2.5mm}\par\noindent {\em \thesubsection . #1}\par
 \vspace{0.5mm} }
\renewcommand{\thebibliography}[1]{ {\vspace{5mm}\par \noindent{\bf
References}\par \vspace{2mm}}
\list
 {\arabic{enumi}.}{\settowidth\labelwidth{[#1]}\leftmargin\labelwidth
 \advance\leftmargin\labelsep\addtolength{\topsep}{-4em}
 \usecounter{enumi}}
 \def\newblock{\hskip .11em plus .33em minus .07em}
 \sloppy\clubpenalty4000\widowpenalty4000
 \sfcode`\.=1000\relax \setlength{\itemsep}{-0.4em} }
{\hfill{ULB-TH/99-15, VUB/TENA/99/7, hep-th/9909094}}

\vspace{1cm}

\begin{center}
{\bf DEFORMATIONS OF CHIRAL TWO-FORMS IN SIX DIMENSIONS }

\vspace{1.4cm}

XAVIER BEKAERT${}^1$, MARC HENNEAUX${}^1$ and
ALEXANDER SEVRIN${}^2$ 

\vspace{.1cm}

${}^1${\em Physique Th\'eorique et Math\'ematique, Universit\'e Libre de Bruxelles,}\\
{\em Campus Plaine C.P. 231, B-1050 Bruxelles, Belgium} \\
${}^2${\em Theoretische Natuurkunde, Vrije Universiteit Brussel} \\
{\em Pleinlaan 2, B-1050 Brussel, Belgium} \\
\end{center}
\centerline{ABSTRACT}
\vspace{- 4 mm}  
\begin{quote}\small
Motivated by a system consisting of a number of parallel M5-branes,
we study possible local deformations of chiral two-forms in six dimensions.
Working to first order in the coupling constant, this reduces to the study of
the local BRST cohomological group at ghost number zero.
We obtain an exhaustive list of all possible deformations. None of them allows 
for a satisfactory formulation of the M5-branes system leading to the 
conclusion that no local field theory can describe such a system.
\end{quote}
\baselineskip18pt
\addtocounter{section}{1}
\noindent

\vspace{5mm}

The M5-brane is perhaps the most elusive object in M-theory \cite{sw}. In the limit 
where bulk gravity decouples, it is described by a six-dimensional field 
theory. In order to match the eight propagating fermionic degrees of  freedom,
its bosonic sector has to include, besides the five scalar fields  which
describe the position of the brane in transverse space, a chiral  two-form
transforming as the (3,1) of the little group $SU(2)\times  SU(2)$. The latter
reflects that M2-branes may end on M5-branes. The  resulting theory is an
$N=(2,0)$ superconformal field theory in six  dimensions \cite{ls}.

A single M5 brane with strong classical fields is well understood;
its Lagrangian is described in \cite{PS,S,APPS,PST,BLNPST}.
However, when several, say $n$, M5-branes coincide, 
little is known. Compactifying one dimension of M-theory on  a circle yields
type IIA string theory. If the M5-branes are taken  transversal to the compact
direction, they become $n$ coinciding  NS5-branes in type IIA, again a poorly
understood system. However, if the  branes are longitudinal to the compact
direction, the M5-branes appear as a set  of coinciding D4-branes which are
quite well understood. Their dynamics is  governed by a five-dimensional $U(n)$
Born-Infeld theory which, ignoring  higher derivative terms, is an ordinary
$U(n)$ non-abelian gauge theory.  Turning back to the eleven dimensional
picture, this suggests that a non-abelian  extension of the chiral two-form
should exist.

However, there are several indications that this is a highly unusual  system.
Both entropy considerations \cite{kleb,gub} and the calculation of the
conformal  anomaly of the partition function \cite{kostas} show that the theory
should have $n^3$ instead of  $n^2$ degrees of freedom. In  \cite{teitel} it
was argued on geometric  grounds, that for $p>1$, non-chiral $p$-forms do not
allow for non-abelian  extensions. In \cite{knaepen}, geometric prejudices were
dropped, and general deformations of non-chiral $p$-forms were classified to
first order in the coupling constant. Though both  known and novel deformations
were discovered, none of them had the  required property that the $p$-form
gauge algebra becomes genuinely  non-abelian.

In the present letter we will specifically focus on deformations of chiral 
two-forms in six dimensions. By construction, these deformations are continously 
connected to the free theory. We will ignore the fermions and the
scalar fields as we
believe that they will not modify our conclusions. In fact this can easily 
be proven for the scalar fields because they are inert under the two-form 
gauge symmetry.

Our starting point is the action of \cite{hen1},
\begin{equation}
S[A^A_{ij}]= \sum_A\int d^6x (B^{Aij} \dot{A}^A_{ij} - B^{Aij}B^A_{ij})
\label{freeaction}
\end{equation}
for a collection of
$N$ free chiral $2$-forms $A^A_{ij}$, 
($i, j, \dots = 1, \dots, 5$), ($A = 1, \dots, N$), where $N$ is arbitrary 
and could e.g. be equal to $n^3$.  The magnetic fields $
B^{Aij}$ in (\ref{freeaction}) are defined through
\begin{equation}
B^A_{ij} = \frac{1}{3!} \epsilon_{ijklm} F^{Aklm}, \;
F^A_{ijk} = \partial_i A^A_{jk} + \partial_j A^A_{ki} 
+ \partial_k A^A_{ij}.
\end{equation}
If one varies the action with respect to the $2$-forms $A^A_{ij}$,
one gets as equations of motion
\begin{equation}
\epsilon^{ijklm}\partial_0 \partial_k A^A_{lm} - 2 \partial_k F^{Akij} = 0
\Leftrightarrow
\epsilon^{ijklm} \partial_k(\partial_0 A^A_{lm} - B^A_{lm}) =0
\end{equation}
which imply, assuming the second Betti number of the spatial sections to
vanish,
\begin{equation}
\partial_0 A^A_{lm} - B^A_{lm} =  \partial_l u^A_m
- \partial_m u^A_l
\label{EOM}
\end{equation}
for some arbitrary spatial $1$-forms $u^A_m$.  If one identifies $u^A_m$
with $A^A_{0m}$ (which is pure gauge), one may rewrite the equation (\ref{EOM}) as
\begin{equation}
E^A_{ij} = B^{A}_{ij}
\end{equation}
where the $E$'s are the electric fields,
$E^A_{ij} = F^A_{0ij}$.  Covariantly, this is equivalent to the self-duality
condition
$F^{A}_{\lambda \mu \nu} = \! ^{*} F ^{A}_{\lambda \mu \nu}$. 
By gauge-fixing the gauge freedom of the theory,
\begin{equation}
\delta_\Lambda A^A_{ij} = \partial_{i} \Lambda^A_j -
\partial_j \Lambda^A_i
\label{freegaugeinvariance}
\end{equation}
one may set $u^A_m =0$.  One gets then the equations in the ``temporal
gauge", $\partial_0 A^A_{lm} - B^A_{lm} = 0$.

The action (\ref{freeaction}) may be covariantized by adding appropriate
auxiliary and gauge fields.  One gets in this manner the free action
of \cite{PST}.  Conversely, one may fall back on (\ref{freeaction})
by partly gauge-fixing the PST Lagrangian.  
Thus a consistent deformation of the PST Lagrangian defines a
consistent deformation of (\ref{freeaction}).  Though the
action (\ref{freeaction})
is significantly simpler to handle than the PST Lagrangian, we pay the
price that our analysis is not manifestly Lorentz invariant. Without enforcing
Lorentz invariance, we will already obtain strong constraints on the allowed
deformations. We shall come back to Lorentz invariance at the end
of this letter.

Our strategy for studying the possible local deformations of the
action (\ref{freeaction}) is based on the observation that these
are in bijective correpondence with the local BRST
cohomological group $H^{0,6}(s \vert d)$ \cite{BH}, where $s$ is the
BRST differential acting on the fields, the ghosts, and their
conjugate antifields, $d$ is the ordinary space-time exterior derivative and the upper
indices refer to ghost number and form degree resp.  In the present case, $s$ is
given by 
\begin{equation}
s = \delta + \gamma
\end{equation}
with
\begin{eqnarray}
\delta A^A_{ij}&=&\delta C^A_{i}=\delta \eta^A=0,\\
\delta A^{*Aij}&=&2\partial_k F^{Akij}-\epsilon^{ijklm}
\partial_k \dot{A}^A_{lm},\\
\delta C^{*Ai}&=&\partial_j A^{*Aij},\\
\delta \eta^{*A} &=& \partial_i C^{*Ai}
\end{eqnarray}
and
\begin{eqnarray}
\gamma A^A_{ij}&=&\partial_i C^A_j - \partial_j C^A_i,\\
\gamma C^A_i&=&\partial_i \eta^A,\; \; \gamma \eta^A = 0, \label{yep}\\
\gamma A^{*Aij}&=&\gamma C^{*Ai}=\gamma \eta^{*A}=0.
\end{eqnarray}
The $C^A_{i}$ are the ghosts, the $\eta^A$ are the ghosts of ghosts,
while the $A^{*Aij}$, $C^{*Ai}$ and $\eta^{*A}$ are the antifields.
One verifies that $\delta^2=\gamma^2=\delta\gamma+\gamma\delta=0$.
The cocycle condition defining elements of $H^{0,6}(s \vert d)$ is the
``Wess-Zumino condition" at ghost number zero,
\begin{equation}
sa + db = 0, \; \; \; gh(a) = 0
\label{WZ}
\end{equation}
Any solution of (\ref{WZ}) defines a consistent deformation
of the action  (\ref{freeaction}) through
$S[A^A_{ij}] \rightarrow S[A^A_{ij}] + g \int d^6x a_0$,
where $a_0$ is the antifield-independent component of $a$.
The deformation is consistent to first-order in $g$,
in the sense that one can simultaneously deform the
original gauge symmetry (\ref{freegaugeinvariance}) in such
a way that the deformed action is invariant under the
deformed gauge symmetry up to terms of order $g$ (included).  The
antifield-dependent components of $a$ contain informations about
the deformation of the gauge symmetry.  
Trivial solutions of (\ref{WZ}) are of the form $a = \gamma c + de$ and
correspond to $a_0$'s that can be redefined away through field
redefinitions.  Of course, there are also consistency conditions
on the deformations arising from higher-order terms ($g^2$ and
higher), but it turns out that in the case at hand, consistency
to first order already restricts dramatically the possibilities.

There are three possible types of consistent deformations of the action.
First, one may deform the action without
modifying the gauge symmetry.  In that case, $a$ does not
depend on the antifields, $a = a_0$.  These deformations
contain only strictly gauge-invariant terms, i.e., polynomials in
the abelian curvatures and their derivatives (Born-Infeld
terms are in this category) as well as Chern-Simons terms, which are
(off-shell) gauge-invariant under the abelian gauge 
symmetry up to a total derivative. An example of a Chern-Simons term
is given by the kinetic term of (\ref{freeaction}), which can be rewritten 
as $F \wedge \partial_0 A$ (in writing Chern-Simons terms, the
spatial $2$-forms $A^A$ and their successive time derivatives, which are
also spatial $2$-forms, are effectively independent).
Second, one may deform the action and the gauge transformations while
keeping their algebra invariant.  In BRST terms, the corresponding cocycles
involve (non trivially) the antifields $A^{*Aij}$ but not $C^{*Ai}$ or $\eta^{*A}$.
Finally, one may deform everything, including the gauge algebra; the
corresponding cocycles involve all the antifields.

Reformulating the problem of deforming the free action (\ref{freeaction})
in terms of BRST
cohomology enables one to use the powerful tools  of homological
algebra.  Following the approach of \cite{knaepen}, we have 
completely worked out the BRST cohomogical classes at ghost number zero.
In particular, we have established that one can always get rid of the antifields
by adding trivial solutions.  {\it In other words, the
only consistent interactions for a system of chiral $2$-forms in six dimensions
are (up to
redefinitions) deformations that do not 
modify the gauge symmetries (\ref{freegaugeinvariance})
of the free theory.}  These involve the abelian curvatures or Chern-Simons terms.
There are no other consistent, local, deformations. 

We shall give the detailed proof of this assertion
in a separate publication \cite{BHS}.  We shall just outline here the general
skeleton of the proof, which parallels
the analysis of \cite{BBH,knaepen} rather closely, emphasizing only the new features.

To find the general solution of (\ref{WZ}), one expands $a$ according
to the antifields, i.e., more precisely, according to the antighost number,
\begin{equation}
a = a_0 + a_1 + \cdots + a_k, \; \; antigh(a_i) = i.
\label{expansion}
\end{equation}
The only variables with non-vanishing antighost number are the antifields,
with $antigh(A^{*Aij}) = 1$, $antigh (C^{*Ai}) = 2$ and
$antigh(\eta^{*A}) = 3$. A similar expansion holds for $b$.
The fact that $k$ remains finite follows from demanding locality in the 
sense that the number of derivatives in both the deformations of the 
action and in the deformations of the gauge transformations remains 
finite \cite{BBH}.
What we must show is that one can eliminate all the terms in 
(\ref{expansion}) but the antifield-independent component $a_0$.
So, let us assume $k>0$ and finite. 

The last term in the expansion (\ref{expansion})
must fulfill $\gamma a_k + db_k = 0$ from (\ref{WZ}).  As in
the non-chiral case,
one may assume $b_k= 0$ through redefinitions ($k>0$).  Thus,
$\gamma a_k = 0$ and one must
determine the general cocycle of the $\gamma$-differential.
It is here that there is a difference with the non-chiral case.
Indeed,  the time derivatives of the ghosts of ghosts
$\eta^A$ are now in the $\gamma$-cohomology, while they
are trivial in the non-chiral case, where one has  
$\gamma C^A_{0} = \partial_0 \eta^A$.  In the chiral case, however,
there is no ghost $C^{A}_0$, so $\partial_0 \eta^A$ is a non-trivial
$\gamma$-cocycle (at ghost number two).
A similar property holds for the higher-order time derivatives of $\eta^A$.
One easily verifies that these are the only generators of the $\gamma$-cohomology
at positive ghost number.  The other generators of the cohomology
are the curvatures, the antifields and their spacetime derivatives.
Thus, up to trivial terms that
can be absorbed, one may write the last term $a_k$ in (\ref{expansion})
as
\begin{equation}
a_k = \sum_I P^I \omega^I
\label{PI}
\end{equation}
where (i) the $P^I$ are $6$-forms constructed out of the antifields,
the curvatures, their spacetime derivatives and the $dx^\mu$'s; and (ii) the
$\omega^I$ are a basis of the vector space of polynomials in the
ghosts of ghosts $\eta^A$ and their successive time-derivatives.
Furthemore, $antigh(P^I) = antigh (a_k) = k$ (by assumption) and
$gh(\omega^I) = k$ so that $gh(a_k) =gh(\omega^I) - antigh(P^I) = 0$.
This shows that $k$ must be even since the $\eta^A$'s 
and their successive time-derivatives have even ghost number.  Thus, 
ik $k$ is odd, $a_k$ is trivial and can be entirely removed.

Turn now to the next equation following from (\ref{WZ}),
\begin{equation}
\gamma a_{k-1} + \delta a_k + d b_{k-1} = 0
\end{equation}
By following the same line of thought as for non-chiral systems
\cite{BBH}, and using
in addition an argument based on counting time-derivatives of
the ghosts of ghosts, one easily proves that $P^I$
must take the form $P^I = Q^I dx^0$ (up to trivial terms),
where $Q^I$ is a spatial $5$-form (polynomial of degree $5$
in the spatial $dx^k$'s) solution of
$\delta Q^I + \tilde{d} R^I = 0$. Here,
$\tilde{d}$
is the spatial exterior derivative, $d \equiv \partial_k dx^k$.
Furthermore, in order for
(\ref{PI}) to be non-trivial, $Q^I$ must be a non-trivial
solution, i.e., not of the form $\delta M^I + \tilde{d} N^I$.
The analysis of \cite{knaepen} imply that there are non-trivial
solutions of $\delta Q^I + \tilde{d} R^I = 0$ only for
$antigh(Q^I) = 1$ or $3$.  In particular, all solutions
of $\delta Q^I + \tilde{d} R^I = 0$ are trivial in even ghost
number, which is the relevant case for us since
$antigh(Q^I) = antigh (P^I)$.  
There is therefore no way to match the odd antighost number of non-trivial
solutions of $\delta Q^I + \tilde{d} R^I = 0$
with the even ghost number of $\omega^I$ in order to make a non-trivial $a_k$.
Thus, $a_k$ is trivial and can be removed.  The same argument applies 
then to the successive $a_{k-1}$, $a_{k-2}$ ... and we can conclude
that indeed, up to trivial terms, $a$ can be taken not to
depend on the antifields, $a= a_0$.
The only consistent interactions in six
dimensions do not deform the gauge symmetry and are
either strictly gauge-invariant ($\gamma a_0 = 0$), or
gauge-invariant up to a total derivative ($\gamma a_0 +
db_0 = 0$)\footnote{A similar result holds for non-chiral
$2$-forms in six dimensions, for which the only symmety-deforming
consistent interactions are 
the Freedman-Townsend interactions \cite{FT,knaepen}, but these are available
in four dimensions only).}.      

Because they do not deform the gauge-symmetry,
the off-shell gauge-invariant (up to a possible
total derivative) interactions are clearly consistent to all
orders, so there is no further constraint following from 
gauge invariance.  If one imposes in addition Lorentz-invariance,
then, one gets of course additional restrictions on the gauge-invariant
interactions. In the case where the
Lagrangian is required to involve only first-order derivatives
of the fields, these restrictions are most easily analysed by
using the Dirac-Schwinger criterion, which easily leads to
the Perry-Schwarz condition \cite{PS} on the
Hamiltonian in the case of a single
field \cite{bekaert}.  We have, however, not
done it explicitly for a system with many
chiral $2$-forms\footnote{The Dirac-Schwinger criterion
is also useful for the related problem of
manifestly duality-invariant formulations of electromagnetism
in $4$ dimensions \cite{deser} and reproduces 
there the condition of \cite{GR}.}.  
The Dirac-Schwinger criterion also implies
consistent gravitational coupling of the chiral $2$-forms
\cite{hen1,bekaert}.
 
The present analysis clearly leads to the conclusion that all continous, local
deformations yield abelian 
algebras. In other words, {\it no local field theory of chiral two-forms continously 
connected with the free theory, can describe 
a system of $n$ coinciding $M5$-branes}. This leaves of course the 
non-local deformations of the abelian theory. Proposals in this direction 
exist \cite{nepo,orlando} where the two-form is used to construct a 
connection on the principal bundle based on the space of loops with a common point. 
However, this approach requires the introduction of a one-form potential which 
is used to parallel transport the two-form from the common point
to some point on the loop. Such a 
one-form potential doesn't seem to appear in the 
M5-system. A way out would be to constrain the potentials to be flat, 
but even then one finds that also here the algebra remains an abelian one.
Finally, three-form field-strengths and their two-form potentials find a natural geometrical
setting in the context of gerbes \cite{gerbe}. However, there as well, a non-abelian 
extension of the two-form gauge-symmetry is still lacking.

\vspace{5mm}

\noindent {\bf Acknowledgments}:
We thank Kostas Skenderis and Jan Troost for dicussions.
X.B. and M.H. are supported in part by the ``Actions de
Recherche Concert{\'e}es" of the ``Direction de la Recherche
Scientifique - Communaut{\'e} Fran{\c c}aise de Belgique", by
IISN - Belgium (convention 4.4505.86) and by
Proyectos FONDECYT 1970151 and 7960001 (Chile). 
A.S. is supported in part by the FWO and by the European
Commission TMR programme ERBFMRX-CT96-0045 in which he is
associated to K.\ U.\ Leuven.


\begin{thebibliography}{99}
\bibitem{sw} A. Strominger, Phys. Lett. {\bf B383} (1996) 44, {\tt hep-th/9512059}; 
E. Witten, J. Geom. Phys. {\bf 22} (1997) 103, {\tt hep-th/9610234};
E. Witten, JHEP {\bf 9801} (1998) 001, {\tt hep-th/9710065}
\bibitem{ls} N. Seiberg, Phys. Lett. {\bf B408} (1997) 98, {\tt hep-th/9705221};
M. Berkooz, M. Rozali and N. Seiberg, Phys. Lett. {\bf 408} (1997) 105, {\tt hep-th/9704089}
\bibitem{PS} M. Perry and J. H. Schwarz,
Nucl. Phys. {\bf B489} (1997) 47-64, {\tt hep-th/9611065} 
\bibitem{S} J. H. Schwarz, Phys. Lett. {\bf B395} (1997) 191,
{\tt hep-th/9701008}
\bibitem{APPS} M. Aganagic, J. Park, C. Popescu, J. H. Schwarz,
Nucl. Phys. {\bf B496} (1997) 191, {\tt hep-th/9701166}
\bibitem{PST} P. Pasti, D. Sorokin and M. Tonin,
Phys. Lett. {\bf B398} (1997) 41, {\tt hep-th/9701037}
\bibitem{BLNPST} I. Bandos, K. Lechner, A. Nurmagambetov,
P. Pasti, D. Sorokin and M. Tonin, Phys. Rev. Lett. {\bf 78} (1997) 4332,
{\tt hep-th/9701149}
\bibitem{kleb} I.R. Klebanov and A.A. Tseytlin, Nucl. Phys. {\bf B475} 
(1996) 164, {\tt hep-th/9604089}
\bibitem{gub} S.S. Gubser and I.R. Klebanov, Phys. Lett. {\bf B413} 41, 
{\tt hep-th/9708005}
\bibitem{kostas} M. Henningson and K. Skenderis, JHEP {\bf 9807} (1998) 023, 
{\tt hep-th/9806087}
\bibitem{teitel} C. Teitelboim, Phys. Lett. {\bf 167B} (1986) 63
\bibitem{knaepen} M. Henneaux and B. Knaepen, Phys. Rev. {\bf D56} (1997) 
6076, {\tt hep-th/9706119}; M. Henneaux, Phys. Lett. {\bf B368} (1996)
83, {\tt hep-th/9511145};
M. Henneaux and B. Knaepen, Nucl. Phys. {\bf B548} (1999) 491,
{\tt hep-th/9812140};
M. Henneaux, B. Knaepen and C. Schomblond,  Comm .Math. Phys. 
{\bf 186} (1997) 137,
{\tt hep-th/9606181}
\bibitem{hen1} M. Henneaux and C. Teitelboim, Phys. Lett. {\bf B206} 
(1988) 650
\bibitem{BH} G. Barnich and M. Henneaux, Phys.Lett. {\bf B311} (1993) 123,
{\tt hep-th/9304057}
\bibitem{BHS} X. Bekaert, M. Henneaux and A. Sevrin, in preparation
\bibitem{BBH} G. Barnich, F. Brandt and M. Henneaux,
Comm. Math. Phys {\bf 174} (1995) 93, {\tt hep-th/9405194 }
\bibitem{FT} D. Z. Freedman and P. K. Townsend, Nucl.
Phys. {\bf B177} (1981) 282
\bibitem{bekaert} X. Bekaert and M. Henneaux, Int. J. Th. Phys.
{\bf 38} (1999) 1161, {\tt hep-th/9806062}
\bibitem{deser} S. Deser and O. Sarioglu, Phys. Lett. {\bf B423}
(1998) 369, {\tt hep-th/9712067}
\bibitem{GR} G. W. Gibbons and D. A. Rasheed, Nucl. Phys.
{\bf B454} (1995) 185,
{\tt hep-th/9506035}
\bibitem{nepo} R.I. Nepomechie, Nucl. Phys. {\bf B212} (1983) 301
\bibitem{orlando} O. Alvarez, L.A. Ferreira and J. Sanchez-Guillen, Nucl. 
Phys. {\bf B529} (1998) 689, {\tt hep-th/9710147} 
\bibitem{gerbe} J. Kalkkinen, JHEP {\bf 9907} (1999) 002, {\tt hep-th/9905018}; 
N. Hitchin, {\tt math.DG/9907034}
\end{thebibliography}
\end{document}